\documentclass[aps,prl,twocolumn,floatfix,10pt,amssymb,amsfont,amsmath,superscriptaddress,float,longbibliography,nobalancelastpage,nofootinbib]{revtex4-2}

\usepackage{graphicx}
\usepackage{float}
\usepackage{tikz}
\usetikzlibrary{shapes.geometric,arrows,matrix,calc,scopes,decorations.markings,intersections,arrows.meta,decorations.pathreplacing,angles,quotes,patterns,decorations.pathmorphing}
\usepackage{ifthen}
\usepackage{amsmath}
\usepackage{braket}
\allowdisplaybreaks

\tikzset{->-/.style={decoration={
			markings,
			mark=at position #1 with {\arrow{>}}},postaction={decorate}}
}

\tikzset{-<-/.style={decoration={
			markings,
			mark=at position #1 with {\arrow{<}}},postaction={decorate}}
}

\newcommand{\sss}{\scriptstyle}

\newcommand{\circleNode}[3]{
\draw [fill = white] (#1,#2) circle (.33);
\node at (#1,#2) {${\sss #3}$};
}

\newcommand{\circleNodeShaded}[3]{
\draw [preaction={fill=white}, pattern = north east lines, pattern color = gray!60] (#1,#2) circle (.33);
\node at (#1,#2) {${\sss #3}$};
}

\newcommand{\triangleRightNodeShaded}[3]{
\draw [preaction={fill=white}, pattern = north east lines, pattern color = gray!60] ($(#1,#2)+(0.2,0.3)$) -- ($(#1,#2)+(-0.3,0.3)$) -- ($(#1,#2)+(-0.3,-0.3)$) -- ($(#1,#2)+(0.2,-0.3)$) -- ($(#1,#2)+(0.3,0)$) -- cycle;
\node at (#1,#2) {${\sss #3}$};
}

\newcommand{\triangleRightNode}[3]{
\draw [fill = white] ($(#1,#2)+(0.2,0.3)$) -- ($(#1,#2)+(-0.3,0.3)$) -- ($(#1,#2)+(-0.3,-0.3)$) -- ($(#1,#2)+(0.2,-0.3)$) -- ($(#1,#2)+(0.3,0)$) -- cycle;
\node at (#1,#2) {${\sss #3}$};
}

\newcommand{\triangleUpNodeShaded}[3]{
\draw [preaction={fill=white}, pattern = north east lines, pattern color = gray!60] ($(#1,#2)+(-0.3,0.2)$) -- ($(#1,#2)+(-0.3,-0.3)$) -- ($(#1,#2)+(0.3,-0.3)$) -- ($(#1,#2)+(0.3,0.2)$) -- ($(#1,#2)+(0,0.3)$) -- cycle;
\node at (#1,#2) {${\sss #3}$};
}

\newcommand{\triangleUpNode}[3]{
\draw [fill = white] ($(#1,#2)+(-0.3,0.2)$) -- ($(#1,#2)+(-0.3,-0.3)$) -- ($(#1,#2)+(0.3,-0.3)$) -- ($(#1,#2)+(0.3,0.2)$) -- ($(#1,#2)+(0,0.3)$) -- cycle;
\node at (#1,#2) {${\sss #3}$};
}

\newcommand{\triangleDownNode}[3]{
\draw [fill = white] ($(#1,#2)+(-0.3,-0.2)$) -- ($(#1,#2)+(-0.3,0.3)$) -- ($(#1,#2)+(0.3,0.3)$) -- ($(#1,#2)+(0.3,-0.2)$) -- ($(#1,#2)+(0,-0.3)$) -- cycle;
\node at (#1,#2) {${\sss #3}$};
}

\newcommand{\squareNode}[3]{
\draw [fill = white] ($(#1,#2)+(0.3,0.3)$) -- ($(#1,#2)+(-0.3,0.3)$) -- ($(#1,#2)+(-0.3,-0.3)$) -- ($(#1,#2)+(0.3,-0.3)$) -- cycle;
\node at (#1,#2) {${\sss #3}$};
}

\newcommand{\FixedPointEq}[1]{
\begin{tikzpicture}[baseline={([yshift=-.55ex]current bounding box.center)}]
\ifthenelse{#1=1}{
\draw (0,-1) -- (0,1);
\draw (-1,0) -- (.75,0);
\draw[rounded corners = 5pt] (-1,0) -- (-1,1) -- (0,1);
\draw[rounded corners = 5pt] (-1,0) -- (-1,-1) -- (0,-1);
\draw (0,1) -- (.75,1);
\draw (0,-1) -- (.75,-1);
\circleNode{0}{0}{T};
\triangleRightNode{0}{1}{A_L};
\triangleRightNode{0}{-1}{A_L};
\squareNode{-1}{0}{A_s};
}{}
\ifthenelse{#1=2}{
\draw (-1,0) -- (-.25,0);
\draw[rounded corners = 5pt] (-1,0) -- (-1,1) -- (-.25,1);
\draw[rounded corners = 5pt] (-1,0) -- (-1,-1) -- (-.25,-1);
\squareNode{-1}{0}{A_s};
}{}
\ifthenelse{#1=3}{
\draw (-.75,0) -- (0,0);
\draw (0,0) -- (.75,0);
\draw (0,0) -- (0,-.75);
\squareNode{0}{0}{A_s};
}{}
\ifthenelse{#1=4}{
\draw (-1,0) -- (0,0);
\draw (0,0) -- (1,0);
\draw (1,0) -- (1.75,0);
\draw (0,0) -- (0,-.75);
\draw (-1.75,0) -- (-1,0);
\triangleRightNode{0}{0}{A_L};
\circleNode{-1}{0}{\frac{1}{X}};
\circleNode{1}{0}{X};
}{}
\ifthenelse{#1=5}{
\draw[rounded corners = 5pt] (0,-1) -- (-.75,-1) -- (-.75,0) -- (0,0);
\draw (0,0) -- (0,-1);
\draw (0,0) -- (.75,0);
\draw (0,-1) -- (.75,-1);
\triangleRightNode{0}{0}{A_L};
\triangleRightNode{0}{-1}{\bar{A}_L}
}{}
\ifthenelse{#1=6}{
\draw[rounded corners = 5pt] (0,-1) -- (-.5,-1) -- (-.5,0) -- (0,0);
}{}
\end{tikzpicture}
}
\newcommand{\AlternativeFixedPointEq}[1]{
\begin{tikzpicture}[baseline={([yshift=-.55ex]current bounding box.center)}]
\ifthenelse{#1=1}{
\draw (0,-1) -- (0,1);
\draw (-1,0) -- (1.55,0);
\draw (1,1) -- (1.55,1);
\draw (1,-1) -- (1.55,-1);
\draw (-1, 1) -- (0,1);
\draw (0,1) -- (1,1);
\draw (-1, -1) -- (-1,0);
\draw (-1,0) -- (-1,1);
\draw (-1, -1) -- (0,-1);
\draw (0,-1) -- (1,-1);
\circleNode{0}{0}{T};
\triangleRightNode{0}{1}{A_L};
\triangleUpNode{-1}{0}{A_L};
\triangleRightNode{0}{-1}{A_L};
\circleNode{-1}{1}{X};
\circleNode{-1}{-1}{\frac{1}{X}};
\circleNode{1}{1}{X};
\circleNode{1}{-1}{X};
}{}

\ifthenelse{#1=2}{
\draw (-1,1) -- (-.25,1);
\draw (-1,0) -- (-.25,0);
\draw (-1,0) -- (-1,1);
\draw[rounded corners = 5pt] (-1,0) -- (-1,-1) -- (-.25,-1);
\circleNode{-1}{1}{X^2};
\triangleUpNode{-1}{0}{A_L};

\phantom{\circleNode{-1}{-1}{}};
}{}
\end{tikzpicture}
}

\newcommand{\DerivativeFixedPointEq}[1]{
\begin{tikzpicture}[
baseline={([yshift=-.55ex]current bounding box.center)}]
\ifthenelse{#1=1}{
\draw (0,-1) -- (0,1);
\draw (-1,0) -- (1.55,0);
\draw (1,1) -- (1.55,1);
\draw (1,-1) -- (1.55,-1);
\draw (-1, 1) -- (0,1);
\draw (0,1) -- (1,1);
\draw (-1, -1) -- (-1,0);
\draw (-1,0) -- (-1,1);
\draw (-1, -1) -- (0,-1);
\draw (0,-1) -- (1,-1);

\circleNodeShaded{0}{0}{dT};

\triangleRightNode{0}{1}{A_L};
\triangleUpNode{-1}{0}{A_L};
\triangleRightNode{0}{-1}{A_L};

\circleNode{-1}{1}{X};
\circleNode{-1}{-1}{\frac{1}{X}};
\circleNode{1}{1}{X};
\circleNode{1}{-1}{X};
}{}

\ifthenelse{#1=2}{
\draw (0,-1) -- (0,1);
\draw (-1,0) -- (1.55,0);
\draw (1,1) -- (1.55,1);
\draw (1,-1) -- (1.55,-1);
\draw (-1, 1) -- (0,1);
\draw (0,1) -- (1,1);
\draw (-1, -1) -- (-1,0);
\draw (-1,0) -- (-1,1);
\draw (-1, -1) -- (0,-1);
\draw (0,-1) -- (1,-1);

\circleNode{0}{0}{T};
\triangleRightNodeShaded{0}{1}{dA_L};
\triangleUpNode{-1}{0}{A_L};
\triangleRightNode{0}{-1}{A_L};

\circleNode{-1}{1}{X};
\circleNode{-1}{-1}{\frac{1}{X}};
\circleNode{1}{1}{X};
\circleNode{1}{-1}{X};
}{}

\ifthenelse{#1=3}{
\draw (0,-1) -- (0,1);
\draw (-1,0) -- (1.55,0);
\draw (1,1) -- (1.55,1);
\draw (1,-1) -- (1.55,-1);
\draw (-1, 1) -- (0,1);
\draw (0,1) -- (1,1);
\draw (-1, -1) -- (-1,0);
\draw (-1,0) -- (-1,1);
\draw (-1, -1) -- (0,-1);
\draw (0,-1) -- (1,-1);

\circleNode{0}{0}{T};
\triangleRightNode{0}{1}{A_L};
\triangleUpNodeShaded{-1}{0}{dA_L};
\triangleRightNode{0}{-1}{A_L};
\circleNode{-1}{1}{X};
\circleNode{-1}{-1}{\frac{1}{X}};
\circleNode{1}{1}{X};
\circleNode{1}{-1}{X};
}{}

\ifthenelse{#1=4}{
\draw (0,-1) -- (0,1);
\draw (-1,0) -- (1.55,0);
\draw (1,1) -- (1.55,1);
\draw (1,-1) -- (1.55,-1);
\draw (-1, 1) -- (0,1);
\draw (0,1) -- (1,1);
\draw (-1, -1) -- (-1,0);
\draw (-1,0) -- (-1,1);
\draw (-1, -1) -- (0,-1);
\draw (0,-1) -- (1,-1);

\circleNode{0}{0}{T};
\triangleRightNode{0}{1}{A_L};
\triangleUpNode{-1}{0}{A_L};
\triangleRightNodeShaded{0}{-1}{dA_L};
\circleNode{-1}{1}{X};
\circleNode{-1}{-1}{\frac{1}{X}};
\circleNode{1}{1}{X};
\circleNode{1}{-1}{X};
}{}

\ifthenelse{#1=5}{
\draw (0,-1) -- (0,1);
\draw (-1,0) -- (1.55,0);
\draw (1,1) -- (1.55,1);
\draw (1,-1) -- (1.55,-1);
\draw (-1, 1) -- (0,1);
\draw (0,1) -- (1,1);
\draw (-1, -1) -- (-1,0);
\draw (-1,0) -- (-1,1);
\draw (-1, -1) -- (0,-1);
\draw (0,-1) -- (1,-1);

\circleNode{0}{0}{T};
\triangleRightNode{0}{1}{A_L};
\triangleUpNode{-1}{0}{A_L};
\triangleRightNode{0}{-1}{A_L};
\circleNodeShaded{-1}{1}{dX};
\circleNode{-1}{-1}{\frac{1}{X}};
\circleNode{1}{1}{X};
\circleNode{1}{-1}{X};
}{}

\ifthenelse{#1=6}{
\draw (0,-1) -- (0,1);
\draw (-1,0) -- (1.55,0);
\draw (1,1) -- (1.55,1);
\draw (1,-1) -- (1.55,-1);
\draw (-1, 1) -- (0,1);
\draw (0,1) -- (1,1);
\draw (-1, -1) -- (-1,0);
\draw (-1,0) -- (-1,1);
\draw (-1, -1) -- (0,-1);
\draw (0,-1) -- (1,-1);

\circleNode{0}{0}{T};
\triangleRightNode{0}{1}{A_L};
\triangleUpNode{-1}{0}{A_L};
\triangleRightNode{0}{-1}{A_L};
\circleNode{-1}{1}{X};
\circleNodeShaded{-1}{-1}{d\frac{1}{X}};
\circleNode{1}{1}{X};
\circleNode{1}{-1}{X};
}{}

\ifthenelse{#1=7}{
\draw (0,-1) -- (0,1);
\draw (-1,0) -- (1.55,0);
\draw (1,1) -- (1.55,1);
\draw (1,-1) -- (1.55,-1);
\draw (-1, 1) -- (0,1);
\draw (0,1) -- (1,1);
\draw (-1, -1) -- (-1,0);
\draw (-1,0) -- (-1,1);
\draw (-1, -1) -- (0,-1);
\draw (0,-1) -- (1,-1);

\circleNode{0}{0}{T};
\triangleRightNode{0}{1}{A_L};
\triangleUpNode{-1}{0}{A_L};
\triangleRightNode{0}{-1}{A_L};
\circleNode{-1}{1}{X};
\circleNode{-1}{-1}{\frac{1}{X}};
\circleNodeShaded{1}{1}{dX};
\circleNode{1}{-1}{X};
}{}

\ifthenelse{#1=8}{
\draw (0,-1) -- (0,1);
\draw (-1,0) -- (1.55,0);
\draw (1,1) -- (1.55,1);
\draw (1,-1) -- (1.55,-1);
\draw (-1, 1) -- (0,1);
\draw (0,1) -- (1,1);
\draw (-1, -1) -- (-1,0);
\draw (-1,0) -- (-1,1);
\draw (-1, -1) -- (0,-1);
\draw (0,-1) -- (1,-1);

\circleNode{0}{0}{T};
\triangleRightNode{0}{1}{A_L};
\triangleUpNode{-1}{0}{A_L};
\triangleRightNode{0}{-1}{A_L};
\circleNode{-1}{1}{X};
\circleNode{-1}{-1}{\frac{1}{X}};
\circleNode{1}{1}{X};
\circleNodeShaded{1}{-1}{dX};
}{}

\ifthenelse{#1=9}{
\draw (-1,1) -- (-.25,1);
\draw (-1,0) -- (-.25,0);
\draw (-1,0) -- (-1,1);
\draw[rounded corners = 5pt] (-1,0) -- (-1,-1) -- (-.25,-1);
\circleNode{-1}{1}{X^2};
\triangleUpNode{-1}{0}{A_L};
\phantom{\circleNode{-1}{-1}{}};
}{}

\ifthenelse{#1=10}{
\draw (-1,1) -- (-.25,1);
\draw (-1,0) -- (-.25,0);
\draw (-1,0) -- (-1,1);
\draw[rounded corners = 5pt] (-1,0) -- (-1,-1) -- (-.25,-1);

\circleNodeShaded{-1}{1}{dX^2};
\triangleUpNode{-1}{0}{A_L};

\phantom{\circleNode{-1}{-1}{}};
}{}

\ifthenelse{#1=11}{
\draw (-1,1) -- (-.25,1);
\draw (-1,0) -- (-.25,0);
\draw (-1,0) -- (-1,1);
\draw[rounded corners = 5pt] (-1,0) -- (-1,-1) -- (-.25,-1);

\circleNode{-1}{1}{X^2};
\triangleUpNodeShaded{-1}{0}{dA_L};
\phantom{\circleNode{-1}{-1}{}};
}{}
\end{tikzpicture}
}

\newcommand{\dALeqn}[1]{
\begin{tikzpicture}[baseline={([yshift=-.55ex]current bounding box.center)}]
\ifthenelse{#1=1}{
\draw (0,0) -- (0,-.75);
\draw (0,0) -- (-.75,0);
\draw (0,0) -- (.75,0);
\draw[decorate, decoration={snake}] (0,0) -- (0.712, 0.712);
\triangleRightNodeShaded{0}{0}{dA_L};
}{}

\ifthenelse{#1=2}{
\draw (-1,0) -- (-1.75,0);
\draw (-1,0) -- (-1,-.75);
\draw[double, double distance=1.5pt] (-1,0) -- (0,0);
\draw[decorate, decoration={snake}] (0,0) -- (0.712, 0.712);
\draw (0,0) -- (.75,0);
\draw (1,0) -- (1.75,0);

\circleNode{-1}{0}{A_{L}^{\perp}};
\circleNodeShaded{0}{0}{dQ};
\circleNode{1}{0}{X^{-2}};

}{}

\end{tikzpicture}
}

\newcommand{\SquareProj}[1]{
\begin{tikzpicture}[baseline={([yshift=-.55ex]current bounding box.center)}]
\ifthenelse{#1=1}{
\draw (0,1) -- (1,1);
\draw (0,0) -- (.75,0);
\draw[double, double distance=1.5pt] (.75,0) -- (.75,.5);
\draw[rounded corners= 2pt, double, double distance=1.5pt] (.75,.5) -- (.75,.75)-- (1,.75);

\draw[rounded corners = 5pt] (.75,0) -- (.75,-1) -- (0,-1);

\draw[pattern=north east lines] (-1,-1.1) rectangle (0,1.1);
\circleNode{-.5}{0}{\tilde{M}};
\triangleUpNode{.75}{0}{A_{L}^{\perp}};
\phantom{\circleNode{.75}{1}{}};
}{}

\ifthenelse{#1=2}{

\draw (0,1) -- (1.75,1);
\draw (0,0) -- (1.75,0);

\draw[pattern=north east lines] (-1,-1.1) rectangle (0,1.1);
\draw[decorate, decoration={snake}] (1.75, 1) -- (2.5,1.75);
\draw  (1.75,1) -- (1.75,0) ;

\draw[rounded corners = 5pt] (1.75,0) -- (1.75,-1) -- (0,-1);

\circleNode{-.5}{0}{\tilde{M}};
\circleNode{.75}{1}{X^2};
\circleNode{1.75}{1}{S};
\triangleUpNode{1.75}{0}{A_{L}};
}{}
\end{tikzpicture}
}

\newcommand{\EuclideanNorm}[1]{
\begin{tikzpicture}[baseline={([yshift=-.55ex]current bounding box.center)}]

\ifthenelse{#1=1}{
\draw (0,-1.5) -- (0,1);
\draw (-1,0) -- (1,0);
\draw (1,1) -- (1,1);
\draw (1,-1) -- (1,-1);

\draw (-1,-1) -- (-1,-1.5);
\draw (-1,-1) -- (-1,0);
\draw (-1,0) -- (-1,1);
\draw (-1,1) -- (0,1);
\draw (0,1) --(1,1);
\draw (1,1) -- (1,0);
\draw (1,0) -- (1,-1);
\draw (1,-1) -- (1,-1.5);

\circleNodeShaded{0}{0}{dT};
\triangleRightNode{0}{1}{A_L};
\triangleUpNode{-1}{0}{A_L};
\triangleDownNode{1}{0}{A_L};
\circleNode{-1}{1}{X};
\circleNode{-1}{-1}{\frac{1}{X}};
\circleNode{1}{1}{X};
\circleNode{1}{-1}{X};
}{}

\end{tikzpicture}
}

\newcommand{\FourNormM}[1]{
\begin{tikzpicture}[baseline={([yshift=-.55ex]current bounding box.center)}]

\ifthenelse{#1=1}{
\draw (0,-1.5) -- (0,1);
\draw (-1,0) -- (1.5,0);
\draw (1,1) -- (1,1);
\draw (1,-1) -- (1,-1);

\draw (-1,-1) -- (-1,-1.5);
\draw (-1,-1) -- (-1,0);
\draw (-1,0) -- (-1,1);
\draw (-1,1) -- (0,1);
\draw (0,1) --(1,1);
\draw (1,1) -- (1.5,1);

\circleNode{0}{0}{T};
\triangleDownNode{-1}{0}{A_L};
\triangleRightNode{0}{1}{A_L};
\circleNode{-1}{-1}{X};
\circleNode{1}{1}{X};
\circleNode{-1}{1}{\frac{1}{X}};

}{}
\ifthenelse{#1=2}{

\draw[rounded corners = 5pt] (1,0)--(1,.75)--(1.5,.75);
\draw (1,0) -- (1.5,0);
\draw (0,-1) --(1,-1);
\draw (1,-1) --(1,0);
\draw (0,-1) -- (0,-1.5);
\draw[rounded corners = 5pt] (0,-1)--(-.75, -1)--(-.75,-1.5);

\triangleDownNode{1}{0}{A_L};
\circleNode{1}{-1}{X};
\triangleRightNode{0}{-1}{A_L};

}{}
\end{tikzpicture}
}

\newcommand{\CornerNormM}[1]{
\begin{tikzpicture}[baseline={([yshift=-.55ex]current bounding box.center)}]

\ifthenelse{#1=1}{
\draw (0,-.75) -- (0,1);
\draw (-1,0) -- (.75,0);
\draw (1,1) -- (1,1);
\draw (1,-1) -- (1,-1);

\draw (-1,-.75) -- (-1,0);
\draw (-1,0) -- (-1,1);
\draw (-1,1) -- (0,1);
\draw (0,1) --(.75,1);

\circleNode{0}{0}{T};
\triangleUpNode{-1}{0}{A_L};
\triangleRightNode{0}{1}{A_L};
\circleNode{-1}{1}{X};

}{}

\end{tikzpicture}
}

\newcommand{\CorrelDerivative}[1]{
\begin{tikzpicture}[baseline={([yshift=-.55ex]current bounding box.center)}]

\ifthenelse{#1=1}{
\draw (0,-1) -- (0,0);

\draw (-1,0) -- (0,0);
\draw (0,0) -- (1,0);
\draw (-1,-1) -- (0,-1);
\draw (0,-1) -- (1,-1);

\triangleRightNodeShaded{0}{0}{dA_L};
\triangleRightNode{0}{-1}{A_L};

}{}

\end{tikzpicture}
}

\newcommand{\PullThrougDerivative}[1]{
\begin{tikzpicture}[baseline={([yshift=-.55ex]current bounding box.center)}]

\ifthenelse{#1=1}{
\node[opacity=0] at (2,-.95) {};
\draw (-.75,0.25) -- (-.25,0.25);
\draw (-.75,0) -- (-.25,0);
\draw (-.75,-0.25) -- (-.25,-0.25);

\draw (.75,0.07) -- (1.5,0.07);
\draw (.75,0) -- (1.5,0);
\draw (.75,-0.07) -- (1.5,-0.07);

\draw[] (-.25, -.75) rectangle (.75, .75);
\node at (0.25,0) {$M_{dA_L}$};

\draw[decorate, decoration={snake}] (1.75,0) -- (2,0.75);
\circleNodeShaded{1.5}{0}{\tilde{dA_L}};
\circleNode{2}{0.75}{x_{dA}};
}{}

\ifthenelse{#1=2}{
\node[opacity=0] at (2,-.95) {};
\draw (-.75,0.25) -- (-.25,0.25);
\draw (-.75,0) -- (-.25,0);
\draw (-.75,-0.25) -- (-.25,-0.25);

\draw[double, double distance=1.5pt] (.75,0) -- (1.5,0);

\draw[] (-.25, -.75) rectangle (.75, .75);
\node at (0.25,0) {$M_{dX}$};

\draw[decorate, decoration={snake}] (1.75,0) -- (2,0.75);
\circleNodeShaded{1.5}{0}{\tilde{dX}};
\circleNode{2}{0.75}{x_{dX}};

}{}

\ifthenelse{#1=3}{
\node[opacity=0] at (2,-.95) {};
\draw (-.75,0.25) -- (-.25,0.25);
\draw (-.75,0) -- (-.25,0);
\draw (-.75,-0.25) -- (-.25,-0.25);

\draw(.75,0) -- (1.5,0);

\draw[] (-.25, -.75) rectangle (.75, .75);
\node at (0.25,0) {$M_{d\lambda}$};

\draw[decorate, decoration={snake}] (1.75,0) -- (2,0.75);
\circleNodeShaded{1.5}{0}{\tilde{d\lambda}};
\circleNode{2}{0.75}{x_{d\lambda}};
}{}

\ifthenelse{#1=4}{
\node[opacity=0] at (2,-.95) {};
\draw (-.75,0.25) -- (-.25,0.25);
\draw (-.75,0) -- (-.25,0);
\draw (-.75,-0.25) -- (-.25,-0.25);

\draw (.75,0.105) -- (1.5,0.105);
\draw (.75,0.035) -- (1.5,0.035);
\draw (.75,-0.035) -- (1.5,-0.035);
\draw (.75,-0.105) -- (1.5,-0.105);

\draw[] (-.25, -.75) rectangle (.75, .75);
\node at (0.25,0) {$M_{dT}$};

\draw[decorate, decoration={snake}] (1.75,0) -- (2,0.75);
\circleNodeShaded{1.5}{0}{\tilde{dT}};
\circleNode{2}{0.75}{b_{dT}};

}{}

\end{tikzpicture}
}

\begin{document}

\title{On the origin of finite entanglement scaling}

\author{Luke Hodgkiss}
\affiliation{Department of Applied Mathematics and Theoretical Physics, University of Cambridge,\\ Wilberforce Road, Cambridge, CB3 0WA, United Kingdom}
\author{Laurens Lootens}
\affiliation{Department of Applied Mathematics and Theoretical Physics, University of Cambridge,\\ Wilberforce Road, Cambridge, CB3 0WA, United Kingdom}
\author{\\Atsushi Ueda}
\affiliation{Department of Physics and Astronomy, Ghent University, Krijgslaan 281, 9000 Gent, Belgium}
\author{Bram Vanhecke}
\affiliation{Department of Physics and Astronomy, Ghent University, Krijgslaan 281, 9000 Gent, Belgium}
\author{Frank Verstraete}
\affiliation{Department of Applied Mathematics and Theoretical Physics, University of Cambridge,\\ Wilberforce Road, Cambridge, CB3 0WA, United Kingdom}
\affiliation{Department of Physics and Astronomy, Ghent University, Krijgslaan 281, 9000 Gent, Belgium}

\begin{abstract}
The concept of finite entanglement scaling forms one of the pillars on which the tensor network ecosystem is built. In this paper, we resolve the open problem of determining the actual perturbations induced by matrix product state approximations of critical systems,  and we demonstrate that these can be quite different than the ones predicted by conformal field theory. To that aim, we develop a sparse linear solver to calculate the forward and backward derivatives of 2-dimensional tensor networks with respect to their defining parameters in an implicit way. This algorithm is of independent interest as it provides a primitive for the variational optimization of projected entangled pair states that circumvents the instabilities plaguing traditional automatic differentiation methods. 
\end{abstract}

\maketitle

Matrix product states (MPS) yield exponentially compressed representations of ground states of one-dimensional quantum Hamiltonians and of leading eigenvectors for transfer matrices of two-dimensional partition functions, and they come equipped with powerful variational optimization algorithms \cite{schollwock2011density,Haegeman_2017}. For higher-dimensional generalizations of MPS known as projected entangled pair states (PEPS) \cite{verstraete2004renormalizationalgorithmsquantummanybody}, however, developing robust optimization algorithms is still an important open problem. The contraction of a PEPS can be reduced by the computation of an MPS approximation to the boundary of a 2d tensor network, and state-of-the-art algorithms proceed by calculating the (backward) derivatives of this MPS with respect to the variational parameters in the PEPS \cite{Vanderstraeten,Corboz,Liao,Naumann}. These derivatives are typically computed using automatic differentiation, but this serves as a black box that is unable to reveal any underlying structures, comes at a significant computational memory cost, and is prone to instabilities.

In this work, we provide a new algorithm for explicitly computing these derivatives without relying on automatic differentiation but rather on implicit differentiation techniques, and we gain new insights into boundary problems for 2d tensor networks. In particular, we are able to answer two questions regarding translation invariant infinite 2d tensor networks. Firstly, which perturbation to the defining tensor leads to the biggest increase or decrease in correlation length? Physically, this amounts to the perturbation of the model which drives it to an RG fixed point, either a gapless critical point or a gapped fixed point respectively. This opens up the path towards a purely MPS-based method for constructing RG flows, which previously relied on tensor network renormalization (TNR) techniques \cite{Evenbly2016LocalScale,PhysRevLett.118.110504,PhysRevLett.118.250602,vanthilt2026practicalintroductiontensornetwork}.

The central question of this paper concerns the approximation of the boundary vector of a critical model with an MPS at finite bond dimension. Working with finite bond dimensions induces a relevant perturbation to the model that opens up a gap, similar to a perturbation induced by finite system size. In analogy to finite size scaling in lattice approximations of conformal field theory(CFTs) \cite{Belavin:1984vu,cardy1996scaling}, this fact has been exploited to construct a theory of finite entanglement scaling \cite{Nishino1996,Tagliacozzo2007,Pollmann2009,Pirvu2012,Zauner2015,Rams2018,Vanhecke2019,Schneider:2024ipg}. The fact that MPS always exhibit a finite correlation length can therefore be seen as feature as opposed to a bug, and it enables the construction of state of the art methods for extracting universal critical properties. This raises an obvious question which has not been previously addressed: for a given critical model and an MPS approximation of fixed bond dimension to its boundary vector, what is this relevant perturbation? We argue that this boils down to finding the perturbation to the defining tensor that provides the optimal improvement to the fidelity of the MPS approximation of the boundary vector of the perturbed model, at a fixed bond dimension. We identify the appropriate Schatten 4-norm for this problem, and show that tools for calculating the backward derivative solve this optimization problem.

Before delving into the technical details, we illustrate our results with the Ising model. At the end of the paper, we will apply our ideas to the problem of determining the relevant perturbation induced by MPS approximation of two decoupled Ising models. Along the way, we derive several useful new technical results concerning the structure and conditioning of corner transfer matrix approximations, and how these can be used for the implicit differentiation of tensor network contractions. We will restrict ourselves to tensor networks on square lattices in which all tensors are real and are invariant under rotations and reflections. An application of the same core principle to the implicit differentiation of the most common two-dimensional tensor network contraction algorithms that do not necessarily assume those symmetries, including CTMRG and boundary MPS, is reported in \cite{burgelman_implicit_2026} and includes benchmarks for PEPS optimizations.
Indeed, combined with a variational pulling through algorithm, the new solver of the backwards derivative can serve as an efficient primitive for state-of-the-art PEPS optimization algorithms without the need for memory-intensive automatic differentiation subroutines.

\section{The incurious case of the Ising model}

As all wisdom in many-body physics starts with the Ising model, we consider the 2D critical classical Ising model on the square lattice. Its partition function can be represented using a 2D tensor network, and its free energy and correlation length can be approximated using pulling-through and CTMRG methods, yielding a series of approximations that converge or scale as a function of the bond dimension $\chi$ of the boundary MPS. Using the tools developed in the next section, we identify the perturbation $T'$ of the local tensor $T$ defined at the critical temperature that induces the biggest change in correlation length.
In Table~\ref{tab:overlaps}, we show the overlap of $T'$ with the perturbation $dT/dh$ obtained by adding a magnetic field $h$ to the Ising model (the spin operator $\sigma$), as a function of the bond dimension $\chi$.

\begin{table}[H]
    \centering
    \begin{tabular}{c|c|c|c|c}
    $\chi$ & 8 & 16 & 32 & 64\\
    \hline
    {\rm $\langle \sigma| T' \rangle$} & 0.99825676 & 0.99987313 & 0.99998821 & 0.99999895
\end{tabular}
    \caption{Overlap of the perturbation $T'$ to the critical Ising tensor that causes the biggest change in correlation length with the spin operator $\sigma$ as a function of bond dimension.}
    \label{tab:overlaps}
\end{table}

This yields a perfect polynomial scaling as a function of $\chi$. The magnetic field perturbation is indeed the most relevant operator in the Ising CFT, matching the intuition that this operator causes the maximal change in correlation length. Note that for smaller bond dimensions, there are small contributions of less relevant terms related to next-nearest neighbour interactions. When optimizing the fidelity of the MPS fixed point as opposed to its correlation length, exactly the same scaling is observed. This seems to justify the central premise of entanglement scaling theory \cite{Nishino1996,Tagliacozzo2007,Pollmann2009,Pirvu2012,Zauner2015,Rams2018,Vanhecke2019} that a finite bond dimension MPS calculation effectively induces the most relevant perturbation predicted by CFT. In contrast to these previous works, however, we are able to determine the nature of this perturbation. We will however see that the simplicity of the Ising model is deceptive, and that the situation is more intricate in general.

Before proceeding to the technical part, let us make some comments. MPS algorithms tend to break symmetry at the critical point, as this leads to a better energy for a given bond dimension. If we repeat the above analysis in the high-temperature phase, the symmetry of the fixed point MPS solution does not leave room for perturbations that break the $\mathbb Z_2$ symmetry, and the most relevant perturbations with respect to both of our measures have a very large overlap with $dT/d\beta$, the energy operator. In this case, there are also notably large overlaps with next-nearest neighbour interactions. Perturbing with the energy operator is clearly less effective in reducing the entanglement than adding a magnetic field, and this explains why simulations at the exact critical point always lead to a symmetry-breaking MPS fixed point - the observed critical temperature always shifts to higher temperatures \cite{Nishino1996}. This idea has been used extensively in actual entanglement scaling protocols, see e.g. \cite{Vanhecke2019}, and has led to the key insight that every quantum spin chain has a dual description that is completely symmetry breaking and whose ground state has a much more efficient description \cite{Lootens1}.

\begin{figure}
    \centering
    \includegraphics[width=0.98\linewidth]{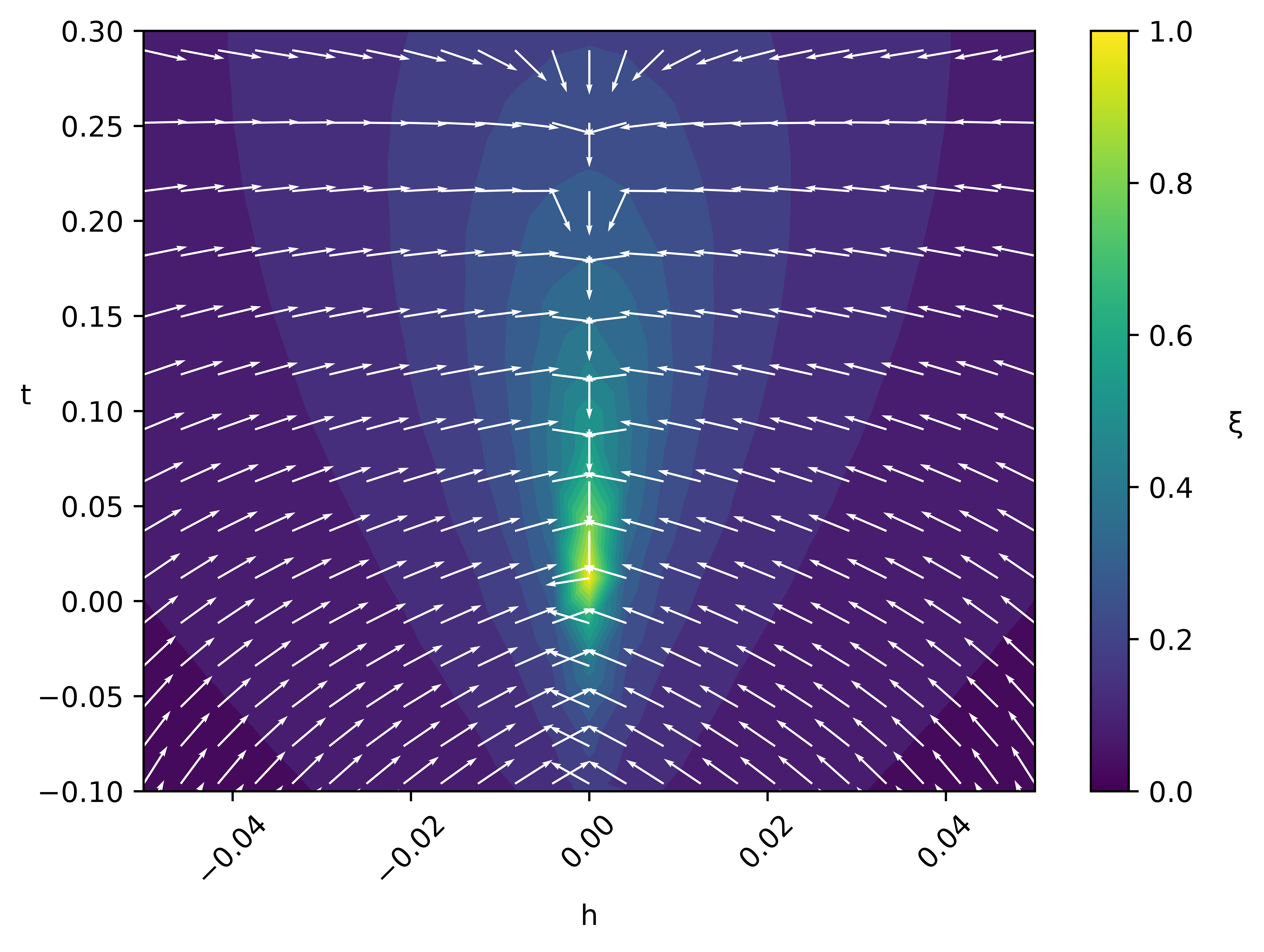}
    \caption{Renormalisation group flows for the 2D classical Ising model at relative temperature $t=\frac{T-T_c}{T_c}$ in a magnetic field $h$. The normalised correlation length $\xi$ is shown in color, and the arrows denote the direction in parameter space that causes the maximal increase in correlation length (calculated by differentiating the correlation length using the implicit method outlined below). These numerical gradients are orthogonal to the iso-correlation length contour lines, verifying the implicit derivative calculation.}
    \label{fig:Ising_phasediagram}
\end{figure}

We conclude our analysis of the 2d classical Ising model by computing the gradient of the correlation length as a function of the temperature and the magnetic field using our implicit differentiation scheme. In Figure~\ref{fig:Ising_phasediagram} this is shown together with the correlation length, demonstrating that our computation of the gradient is correct. This provides a purely MPS based method for constructing RG flow diagrams. 

\section{Implicit derivatives}
For the remainder of this work, we consider a tensor network defined on an infinite  square lattice, and we 
assume that the leading eigenvector of the transfer matrix does not break the $\mathbb D_4$ lattice symmetry (possibly after blocking) and is real. The pulling through algorithm of \cite{Haegeman_2017,Fishman_2018} or standard CTMRG algorithms \cite{nishino1996corner} can be used to identify the optimal finite $\chi$ MPS approximation of this leading eigenvector. In the appropriate gauge, this optimal MPS can be shown to satisfy the following equations:
\begin{align}
    \FixedPointEq{1} \,&=\, \lambda \,\cdot\, \FixedPointEq{2}\,\,, \qquad     \FixedPointEq{5} \,=\, \FixedPointEq{6},\nonumber\\
    \FixedPointEq{3} \,&=\, \FixedPointEq{4}.
\end{align}
Here  $T$ is the tensor of the infinite translational invariant tensor network to be contracted, the MPS tensor $A_L$ is in left canonical form, $A_s$ is the (equivalent) MPS fixed point in the symmetric gauge, and the corner tensor $X$ is real symmetric and satisfies the normalization condition ${\rm Tr}(X^4)=1$. The squares of the (signed) eigenvalues of $X$ are equal to the Schmidt coefficients. Note that we use the notation $d$ for the bond dimension of the tensor $T$ and $\chi$ as the bond dimension of the MPS. Due to the $\mathbb D_4$ symmetry of $T$, it has has $d(d+1)/2+(d-1)d(d+1)(d+2)/8$ linearly independent components. It will be useful to rewrite this equation in a better conditioned variant by embedding it in its natural environment with respect to the 2-norm (more specifically, we want the two-norm of both arguments to be $|\lambda|$):
\begin{align}
    \AlternativeFixedPointEq{1}\,&=\, \lambda \,\cdot\, \AlternativeFixedPointEq{2}.
    \label{eq:FPeq}
\end{align}
This equation enables the determination of the implicit derivative of the  MPS tensors without having to recourse to (unstable) automatic differentiation techniques. Indeed, it completely fixes the optimal MPS $A_L$, $X$ and $\lambda$ -- and therefore it also completely fixes the derivatives $dA_L$, $dX$ and $d\lambda$ as a function of $dT$ (assuming that $dT$ also respects the $\mathbb D_4$ symmetry). The following linear equation relates $dA_L,dX,d\lambda$ with the pertubation $dT$: 
\begin{widetext}
\begin{align*}
&\DerivativeFixedPointEq{1} \,+\, \DerivativeFixedPointEq{2} \,+\, \DerivativeFixedPointEq{3} \,+\, \DerivativeFixedPointEq{4} \,+\,
\DerivativeFixedPointEq{5}\\[1em]
&+\,\,\DerivativeFixedPointEq{6} \,+\, \DerivativeFixedPointEq{7} \,+\, \DerivativeFixedPointEq{8}
\,=\, d\lambda \,\times\, \DerivativeFixedPointEq{9} \,+\, \lambda \,\times\,\DerivativeFixedPointEq{10} \,+\, \lambda \,\times\,\DerivativeFixedPointEq{11}\,\,.
\end{align*}

\end{widetext}
Grouping the different derivatives together, this can be written as a linear problem (where we parametrize a perturbation $dX$ as a linear combination of some set of basis perturbation $\tilde{dX}$ and denote by $x_{dX}$ the expansion of some perturbation $dX$ in this basis):
\begin{align*}
    &\PullThrougDerivative{1} \,+\, \PullThrougDerivative{2}\\[-1em] +\,\, &\PullThrougDerivative{3} \,= \,  \PullThrougDerivative{4}.
\end{align*}
We treat this as a matrix equation: $M \Vec{x} = \Vec{b}$ where $\Vec{x} = (x_{dA_L}, x_{dX}, x_{d\lambda})$ and $\Vec{b}=b_{dT}$. 
To ensure normalization, we also impose that $Tr(X^3 dX)=0$.  Without loss of generality, we can restrict ourselves to derivatives $dA_L$ which keep $A_L$ in left canonical form. By removing the remaining unitary gauge degrees of freedom  and using the natural metric (environment of $A_L$) to make the metric locally Euclidean,  this amounts to parametrising $dA_L$ as:
\begin{align}\label{eq:dA_L param}
    \dALeqn{1} = \dALeqn{2}.
\end{align}
$A_L^\perp$ is a tensor with $\chi^2(d-1)d$ variables and forms a complete left orthogonal basis that is orthogonal to $A_L$, and $dQ$ hence has $\chi^2(d-1)$ degrees of freedom \cite{Vanderstraeten_2019}.

Analogously, we parametrise $dX$ using a basis of Hermitian matrices $dH$  with $\chi (\chi+1)/2$ degrees of freedom:  $X dX+dX X=dH$. In vector notation:
\begin{equation}\label{eq:dX_para}
    |dX\rangle=\frac{I}{X\otimes I+I\otimes X}|dH\rangle
\end{equation}
This preconditioning ensures that $dX$ is Euclidean in its natural environment. Forward derivatives can now be obtained by (sparsely) solving the linear system $\tilde{M}x = dT$ that determines $dA_L,dX$ and $d\lambda$ -- for a total of $(d-1)\chi^2+\chi(\chi+1)/2+1$ degrees of freedom -- as a function of $dT$:
\begin{equation}
    \tilde{M}
    \begin{pmatrix}
        dQ\\
        dX\\
        d\lambda
    \end{pmatrix}
    =\tilde{T}dT
\end{equation}
where $\tilde{M}$ is a matrix with $d\chi^2+1$ rows and $(d-1) \chi^2+\chi (\chi+1)/2+1$ columns. It can easily be verified that $\tilde{T}$ lies in the column space of $\tilde{M}$, and $dT$ is a vector with $d(d+1)/2+(d-1)d(d+1)(d+2)/8-(d(d-1)/2+1)$ components. Note that $d(d-1)/2$ is the number of generators of orthogonal gauge transformations, which we exclude from the allowed gradients together with one extra degree of freedom which amounts to gradients proportional to $T$. The matrix $\tilde{M}$ has more rows than columns, but can be reduced to a square matrix by projecting out its null space:
\begin{align}
    \SquareProj{1} = 0, \quad
    \SquareProj{2}.
\end{align}
Here, $S$ forms a complete basis of $\chi\times\chi$ Hermitian matrices parametrized in such a way that the combination of $X^2,S,A_L$ is an isometry. This gives rise to a linear set of equations with a full-rank square matrix $M'$:
\begin{equation}
    M'
    \begin{pmatrix} 
        dQ\\
        dX\\
        d\lambda
    \end{pmatrix}
    =T'dT.
\end{equation}
This framework can now also be used to calculate backward derivatives, which is the ingredient that is crucial in devising state-of-the-art PEPS optimization algorithms \cite{Vanderstraeten_2016,PhysRevX.9.031041}. Assume that we have a cost function that is written out explicitly as a function \begin{equation}E(T,A_L,X,\lambda)\label{eqE}\end{equation} with $A_L,X,\lambda$ implicit functions of $T$. Then we have 
\begin{align}
\frac{dE}{dT}&=\frac{\partial E}{\partial T}+\frac{\partial E}{\partial A_L}\frac{\partial A_L}{\partial T}+\frac{\partial E}{\partial X}\frac{\partial X}{\partial T}+\frac{\partial E}{\partial \lambda}\frac{\partial \lambda}{\partial T}\\
&=
\frac{\partial E}{\partial T}+
\begin{pmatrix}
\frac{\partial E}{\partial A_L} &\frac{\partial E}{\partial X}& \frac{\partial E}{\partial \lambda}
\end{pmatrix}(M')^{-1}T',
\end{align}
and the problem of calculating the gradient $dT$ thus amounts to  solving the following linear equation:
\begin{equation}
    xM'=
    \begin{pmatrix}
        \frac{\partial E}{\partial A_L} &\frac{\partial E}{\partial X}& \frac{\partial E}{\partial \lambda}
    \end{pmatrix}.
\end{equation}
To make this a well-conditioned problem that can be solved using standard linear solvers such as LSQR and LSMR, it is crucial to use the preconditioning we introduced before. Note that those preconditioners are determined by requiring that all tensors are embedded in their natural environment. We have tested extensively that this preconditioning leads to an $M'$ with a small condition number.

We can associate a Euclidean metric to the perturbations $dT$ by embedding them in their natural environment:
\begin{align}
    \left\lVert \,\, \EuclideanNorm{1} \,\, \right\rVert_2^2.
    \label{fig:Euclides}
\end{align}
This allows for taking overlaps between directions $dT$ in a universal way by taking the (Euclidean) inner product in this Hilbert space of dimension $\chi^2 d$; this is of central importance to get stable PEPS optimization algorithms.

\section{Relevant perturbations induced by $\chi$}
Let us apply this method to the determination of the relevant perturbations induced by finite bond dimension approximations of critical spin systems. 
To that end, we solve the following problem: which infinitesimal change in the defining tensor $T$, while keeping its bond dimension $d$ fixed, results in a better MPS approximation with virtual bond dimension $\chi$.

First of all, we see that an MPS is an exact eigenstate of the transfer matrix if and only if the Schatten 4-norm 
\begin{align}
     \left\lVert \,\, \frac{1}{\lambda} \cdot \FourNormM{1} - \FourNormM{2} \,\, \right\rVert_{4}
     \label{4norm}
\end{align}
is exactly equal to zero. We posit that this 4-norm, with the conditioning chosen in this particular way, is the natural corner transfer matrix analogue of the variance in usual MPS-based algorithms. For any MPS which is not an exact eigenstate, this norm is hence the \emph{natural} norm measuring the quality of the MPS approximation. Defining $Z$ as the matrix
\begin{align}
Z =  \frac{1}{\lambda} \cdot \CornerNormM{1}
\end{align}
where the row indices are labelled by the open vertical legs and the column indices by the open horizontal ones,  the Schatten 4-norm is equal to the fourth root of 
\begin{equation}
    \frac{1}{\lambda^4}{\rm Tr}\left(Z^4\right)-1.
\end{equation}
If we set $E(T,A_l,X,\lambda)$ from Eq.~\eqref{eqE} equal to this nor, we can easily determine the direction $dT$ in which it decreases maximally. This answers the question of determining the perturbation that is induced by working with a finite bond dimension: MPS algorithms effectively induce this perturbation as it leads to the best possible variational MPS approximation.

In a similar vein, we can determine the perturbation of $T$ which induces the largest growth of the correlation length. In that case, the cost function is the second largest eigenvalue of the transfer matrix $E$ of the MPS. Given the left and right eigenvectors $\langle v_L|$ and $|v_R\rangle$ corresponding to the second largest eigenvalue of $E$, this gradient can be obtained from the equation
\begin{align}
    \left( v_L\right) \, \CorrelDerivative{1} \, \left(v_R\right)
\end{align}
It is now straightforward to proceed as before, using backwards derivatives, and determine the perturbation $dT$ which leads to the greatest increase/decrease in correlation length.

\section{The curious case of two Ising models}
In the first section of the paper, we considered the Ising model on the square lattice and calculated the relevant perturbations induced by the finite bond dimension effect. The results were as expected from the usual scaling theory for critical phenomena: the perturbation was the spin operator, the most relevant operator in the Ising CFT. As we will show now, the situation can be more subtle. Let us consider the following problem: what is the relevant perturbation induced by MPS algorithms when simulating a tensor product of two critical systems, in such a way that a single MPS represents the joint leading eigenvector? Even if those two systems are independent, MPS algorithms will entangle the two systems with each other. This allows for keeping more relevant Schmidt coefficients during the truncation process, since the Schmidt coefficients of a tensor product are just the products of the Schmidt coefficients $\lambda_i\lambda_j$, and the smaller ones in this product can therefore be thrown out with minimal cost. This contradicts the intuition from field theory, where this entangling should not occur as the most relevant perturbations do not couple the chains. 

We calculated the perturbation induced by the MPS  by determining the derivative of the Schatten 4-norm in Eq.~\eqref{4norm}. The largest component in this derivative is obtained by a perturbation that amounts to adding a magnetic field to both systems. If we denote $\sigma$ the spins in the first system and $\tau$ in the second, this can be written as the perturbation $\sum_i\sigma^z_i+\tau^z_i$. This is the most relevant perturbation as predicted from conformal field theory. However, the overlap of the two normalized gradients as measured with the natural metric in Eq.~\eqref{fig:Euclides} with the spin operators hovers around $2/3$, even when $\chi$ gets large, which indicates that there are other perturbations that also prevail. The second largest overlap occurs with the global symmetry breaking terms $\sum_{<i,j>}\sigma^z_i\tau^z_i\tau^z_{j}+\sigma^z_i\sigma^z_j\tau^z_i$. This interaction term entangles both spin chains with each other, and it prevails when increasing the bond dimension of the approximating MPS. Note that it is also the most relevant entangling interaction that breaks the diagonal $\mathbb Z_2$ symmetry but preserves the $\mathbb D_4$ lattice symmetry. Finally, the third largest overlap amounts to an on-site Ising interaction between the two systems, of the form $\sum_{<i,j>}\sigma^z_i\tau^z_{i}$. The 9 possible nearest neighbour perturbations in the space of two coupled Ising spin systems account for almost all of the overlap with the gradient, and the other $37$ possible terms appear to be completely irrelevant.

It is interesting to contrast this optimal perturbation with the one that maximizes/minimizes the correlation length. In that case, the optimal gradient amounts to adding a magnetization to only 1 system $\sum_i \sigma^z_i$, with the second most relevant term being then $\sum_{<i,j>}\tau^z_i\sigma^z_j\tau^z_j$. This can be understood from the fact that the optimization breaks the symmetry between the copies and maximally changes the correlation length of one of them, while leaving the other copy unchanged.

Clearly, entanglement scaling theory is not just a simple application of the standard scaling theory of CFT: conformal field theory does not teach us how to obtain the best possible MPS approximation. The current results should serve as a basis for developing a complete theory of MPS scaling.

\section{Discussion and outlook}

We introduced a robust and efficient method for calculating implicit forward and backward derivatives of two-dimensional tensor networks. This method forms a primitive for building robust PEPS optimization algorithms, and our explicit formulation alleviates the uncontrolled discontinuities in gradients that are often encountered in automatic differentiation. 

In the present work, we applied the formalism to the problem of the finite-bond-dimension effect of the MPS. In particular, we determined the effective Hamiltonian whose ground state has the largest overlap with an MPS of a fixed bond dimension $\chi$. For the transverse field Ising chain, this prescription confirmed the conventional picture that the dominant finite-$\chi$ effect is governed by the most relevant perturbation, as discussed in the literature. This simple scenario, however, fails for two decoupled critical Ising chains, where we found non-vanishing contributions from subleading relevant perturbations. This unexpected result motivates a reconsideration of the standard finite-entanglement scaling. Importantly, our method is not limited to the critical case: it can also characterize finite-$\chi$ effects in the gapped system beyond the scope of CFTs. We therefore expect that this formalism will provide a foundation for a more comprehensive theory of finite-bond-dimension effects, both at and away from criticality.  

\section{Acknowledgements}
We are grateful to the authors of \cite{burgelman_implicit_2026} for useful discussions and coordinating our submissions.
L.L. is supported by an EPSRC Postdoctoral Fellowship (grant No. EP/Y020456/1).
FV acknowledges funding from the UKRI grant EP/Z003342/1, BOFGOA
(Grant No. BOF23/GOA/021) and IBOF (Grant No.~IBOF23/064).
AU is supported by the FWO Junior Postdoctoral Fellowship (grant No. 12AHS26N).

\bibliography{bib}

\end{document}